\documentclass[12pt, oneside]{article}

\usepackage{epsfig}
\usepackage{latexsym}
\usepackage{amssymb}
\usepackage{amsmath}
\usepackage{graphicx}
\usepackage{subfigure} 
\usepackage{amsfonts}
\usepackage{array}

\usepackage{fancyhdr}
\fancypagestyle{plain}{
 \fancyhead{} 
 }

\makeatletter
\def\cleardoublepage{\clearpage\if@twoside \ifodd\c@page\else%
	     \hbox{}%
	 \thispagestyle{empty}
	 \newpage%
	 \if@twocolumn\hbox{}\newpage\fi\fi\fi}
\makeatother

\numberwithin{equation}{section}

\newcommand{\be}{\begin{equation}}
\newcommand{\ee}{\end{equation}}
\newcommand{\bea}{\begin{eqnarray}}
\newcommand{\eea}{\end{eqnarray}}

\newcommand{\nn}{\nonumber \\}

\newcommand{\mn}{{\mu\nu}}

\newcommand{\rhod}{{\dot{\rho}}}

\newcommand{\phid}{{\dot{\phi}}}
\newcommand{\rhop}{{\rho^\prime}}
\newcommand{\sigp}{{\sigma^\prime}}
\newcommand{\Np}{{N^\prime}}
\newcommand{\phip}{{\phi^\prime}}
\newcommand{\psip}{{\psi^\prime}}
\newcommand{\psid}{\dot{\psi}}

\newcommand{\Pir}{\Pi_\rho}
\newcommand{\Pis}{\Pi_\psi}
\newcommand{\sub}{\left( \frac{N}{\sigma} \right)}
\newcommand{\gxx}{(g^{11})}
\newcommand{\erho}{e^{2\rho}}
\newcommand{\m}{\mathcal{M}}

\def\gsim{\raise0.3ex\hbox{$>$\kern-0.75em\raise-1.1ex\hbox{$\sim$}}}
\def\lsim{\raise0.3ex\hbox{$<$\kern-0.75em\raise-1.1ex\hbox{$\sim$}}}

\begin{document}

\title{\textbf{Hamiltonian Formulation of Scalar Field Collapse  in Einstein Gauss Bonnet Gravity}}

\author{{\bf T. Taves$^a$, C. D. Leonard$^{b,c}$, G. Kunstatter$^{d}$, R.B. Mann$^{b,e}$ }\\[10pt]
{\small $^a$ Department of Physics and Astronomy and Winnipeg Institute for Theoretical Physics}\\
{\small University of Manitoba, R3T 2N2 Canada}\\ 
{\small   Winnipeg, Manitoba, Canada}\\[5pt] 
{\small $^b$ Department of Physics and Astronomy, University of Waterloo}\\
{\small Waterloo, Ontario, N2L 3G1, Canada}\\
{\small $^c$ Department of Physics and Physical Oceanography}\\
{\small  Memorial University of Newfoundland, St. John's, Newfoundland, A1B 3X7, Canada} \\
{\small $^d$Department of Physics and Winnipeg Institute of Theoretical Physics} \\
{\small University of Winnipeg, Winnipeg, Manitoba, R3B 2E9, Canada} \\
{\small $^e$ Perimeter Institute, 31 Caroline Street North, Waterloo, Ontario, N2L 2Y5, Canada} %
 }
 
 \date{\today}

\maketitle

\begin{abstract}

We compute the Hamiltonian for spherically symmetric scalar field collapse in Einstein-Gauss-Bonnet gravity  in $D$ dimensions using slicings that are regular across future horizons. 
We first reduce the Lagrangian  to two dimensions using spherical symmetry. We then show that choosing the spatial coordinate to be a function of the areal radius leads to a relatively simple Hamiltonian constraint whose gravitational part is the gradient of the generalized mass function.  Next we complete the gauge fixing such that the metric is the Einstein-Gauss-Bonnet generalization of non-static Painlev\'{e}-Gullstrand coordinates.  Finally, we derive the resultant reduced equations of motion for the scalar field.  These equations are suitable for use in numerical simulations of spherically symmetric scalar field collapse in Gauss-Bonnet gravity and can readily be generalized to other matter fields minimally coupled to gravity.

\end{abstract}

\clearpage

\section{Introduction}

The rising popularity of higher dimensional gravitational theories such as string theory calls for the consideration of higher order curvature terms in the action \cite{GBmotiv}.  If we require the action to yield covariant equations that are at most second order in derivatives of the metric, then we must modify the action by adding Lovelock polynomials \cite{Lovelock}, as these are the only known combinations of  higher order curvature terms that satisfy this criterion.  
  These terms are identically zero in four dimensions but are relevant for $D>4$.  The second order Lovelock polynomial is known as the Gauss-Bonnet (GB) term.  
  
A considerable amount is known about static black hole solutions in Lovelock gravity \cite{BH} and there have been several analytic studies of the dynamics of black hole horizons in Einstein-Gauss-Bonnet (EGB) gravity. For example, the effect of  GB terms on the end state of gravitational collapse has been investigated in generalized Vaidya co-ordinates \cite{HM06}. As well, dynamical horizons were studied in the context of EGB gravity in \cite{Booth07} and \cite{HM08a}.  
Surprisingly, comparatively little is known about the formation of black holes in EGB gravity.  This situation stands in strong contrast to that in general relativity, where numerical studies have yielded much information about the formation of black holes from gravitational collapse \cite{collapse}. 
Here we take the first steps in this direction by deriving the Hamiltonian equations describing gravitational collapse of a spherically symmetric scalar field in Einstein-Gauss-Bonnet (EGB) gravity.  


The general form of the Hamiltonian for Lovelock gravity was first derived in \cite{TZ} using a covariant formalism. This work did not write down the explicit form of the Hamiltonian in terms of phase space variables and was therefore not of much use practically for investigating collapse.  Further progress came when the Hamiltonian formulation of vacuum EGB in $D=5$ in terms of Kuchar's geometrodynamic formalism was employed to investigate black hole thermodynamics  \cite{JL97}.  More recently, the role
of surface terms and conserved charges in EGB was explored \cite{AP03,ND04} and 
the mass-energy of a EGB space time was constructed \cite{HM08}.

In the present work, we present a complete Hamiltonian analysis of a massless scalar field  minimally coupled to EGB gravity.   We first derive a partially reduced Hamiltonian in a particular family of gauges in which a function of the area radius is used as the spatial coordinate. This yield slicings that are regular across future (or past) horizons. Moreover, the Hamiltonian constraint can be simply expressed in terms of the generalized mass function first derived in \cite{HM08}. We then fully reduce the theory in generalized Painlev\'e-Gullstrand (PG) coordinates, obtaining the reduced equations of motion for the scalar field and its conjugate. The reduced equations are in a form that is useful for the analysis of gravitational collapse up to and beyond horizon formation either analytically or numerically as done for Einstein gravity in \cite{JZ09}. Our analysis can readily adapted to other types of minimally coupled matter fields.

The paper is organized as follows.  In Section \ref{actionsection} we dimensionally reduce the spherically symmetric, $D$-dimensional action to one space and one time co-ordinate.  We then write the action in a form that manifestly gives second order derivatives of the metric in the covariant equations of motion.  Section \ref{hamiltonian} presents the Hamiltonian analysis and derives the partially gauge fixed equations.  Finally, in Section \ref{gauge} we complete the gauge fixing and calculate Hamilton's equations of motion for a massless scalar field in the presence of gravity in generalized PG co-ordinates.  We then conclude in section \ref{conclusion}. Some technical details are relegated to an Appendix.

\section{The Action}
\label{actionsection}

We start with the Einstein-Hilbert action with a Gauss Bonnet term minimally coupled to a massless scalar field,

\be
S = S_G + S_M = S_{EH} + S_{GB} + S_M,
\ee
where $S_G$ is the gravitational part the action, with contributions from the Einstein-Hilbert term, $S_{EH}$, and the Gauss-Bonnet term, $S_{GB}$ and $S_M$ comes from the scalar field.  These terms are given by

\be
S_{EH} = \frac{1}{16\pi G_D} \int d^{D}x \sqrt{-\tilde{g}} \tilde{R},
\ee

\be
S_{GB} = \alpha \int d^{D}x \sqrt{-\tilde{g}} \left( \tilde{R}^2 - 4\tilde{R}_\mn \tilde{R}^\mn + \tilde{R}_{\mu \nu \rho \sigma}\tilde{R}^{\mu \nu \rho \sigma} \right),
\ee
and

\be
\label{scalaraction}
S_M = -\frac{1}{2}\int d^{D}x \sqrt{-\tilde{g}} |\nabla \psi|^2,
\ee
where $G_D$ is Newton's gravitational constant, $D$ is the number of dimensions, $\tilde{g}$ is the determinant of the metric, $\tilde{R}$ is the Ricci scalar, $\tilde{R}_\mn$ is the Ricci tensor, $\tilde{R}_{\mu \nu \rho \sigma}$ is the Riemann tensor, $\alpha$ is the Gauss Bonnet coupling constant and $\psi$ is the scalar field.

Inserting the spherically symmetric ansatz
\be
ds^2 =  {g}_{ab}(t,x) dx^a dx^b + r^2 \gamma_{ij}dx^i dx^j
\label{4 metric} 
\ee
where $\gamma_{ij}$ is the metric of the unit $n$ sphere, $x$ and $t$ are spatial and time co-ordinates, $r$ is the areal radius and the indices $a$ and $b$ now run from $0$ to $1$.
Performing the angular integrals in the action, we get \cite{HM08} 
\be
S_{EH} = \frac{\nu^{(n)}}{16\pi G_D} \int dxdt \sqrt{-g} r^n \left( R - 2n \frac{\Box r}{r} + n(n-1) \frac{1 - |\nabla r|^2}{r} \right),
\ee
\begin{align}
& S_{GB} = \alpha \nu^{(n)} \int dxdt \sqrt{-g} r^n \frac{4n(n-1)}{r^2} \Biggl[ \frac{1 - |\nabla r|^2}{2}R + (\Box r)^2 - (\nabla_a \nabla_b r)(\nabla^a \nabla^b r) \nn
& + (n-2)(n-3)\frac{[1 - |\nabla r|^2]^2}{4r^2} - (n-2)\frac{1 - |\nabla r|^2}{r} \Box r \Biggr],
\end{align}
and
\be
S_M = -\frac{\nu^{(n)}}{2}\int dxdt \sqrt{-g} r^n |\nabla \psi|^2,
\ee
where $n = D-2$ is the number of angular dimensions and $\nu^{(n)}$ is the surface area of an $n$ sphere. $g$ and $R$ are the determinant and Ricci scalar, respectively, of the 2D metric, $g_{ab}$

We now consider only the gravitational part of the action.  Integrating by parts and defining $\phi = r^{n/2}$   \cite{DM94}, we find 
\begin{align}
S_G & = \nu^{(n)} \int dxdt \sqrt{(-g)} \Biggl[ \left( \alpha 2n(n-1) \phi^{2-4/n} -\alpha \frac{8(n-1)}{n}|\nabla \phi|^2 + \frac{1}{16\pi G_D} \phi^2 \right) R \\ \nonumber
& + \alpha \frac{16(n-1)}{n} \left( \frac{n-2}{n}\left(n^{-1}\phi^{-2}|\nabla \phi|^4 - \phi^{-1}|\nabla \phi|^2 \Box \phi \right) + (\Box \phi)^2 - (\nabla_a \nabla_b \phi)(\nabla^a \nabla^b \phi) \right) \\ \nonumber
& + \alpha 8(n-1)(n-2) \left( n^{-1} \phi^{-4/n} |\nabla \phi|^2 - \phi^{-4/n + 1} \Box \phi +\frac{n(n-3)}{8} \phi^{-8/n + 2} \right) \\ \nonumber
& + \frac{1}{16\pi G_D} n(n-1) \phi^{2-4/n} + \frac{1}{16\pi G_D} \frac{8(n-1)}{n} \frac{1}{2} |\nabla \phi|^2 \Biggr].
\end{align}
A key relationship in two dimensions is the following:
\be
\label{Robb}
\int dxdt \sqrt{(-g)} \left[ -|\nabla \phi|^2 R + 2(\Box \phi)^2 - 2(\nabla_a \nabla_b \phi)(\nabla^a \nabla^b \phi) \right] = 0,
\ee
This allows for further simplification of the action 
\begin{align}
\label{action}
S_G & = \frac{8(n-1)\nu^{(n)}}{16\pi G_D n} \int dxdt \sqrt{(-g)} \Biggl[ \left( \beta 2n(n-1) \phi^{2-4/n} + \frac{n}{8(n-1)} \phi^2 \right) R \\ \nonumber
&-\beta \frac{16(n-1)(n-2)(n-3)}{3n^3} \phi^{-2}|\nabla \phi|^4 \\ \nonumber
& + \beta 8(n-1)(n-2)(n-3) \left( n^{-1} \phi^{-4/n} |\nabla \phi|^2 +\frac{n}{8} \phi^{2 - 8/n} \right) + \frac{n^2}{8} \phi^{2-4/n} + \frac{1}{2} |\nabla \phi|^2 \Biggr],
\end{align}
where $\beta  = 16\pi G_D n \alpha /8(n-1)$ and we have used the identity:
\be
\phi^{-1}|\nabla \phi|^2 \Box \phi = \frac{1}{3}\left(\nabla\cdot(\phi^{-1}|\nabla\phi|^2\nabla\phi) + \phi^{-2}|\nabla \phi|^4 
\right)
\ee  
Equation (\ref{action}) is now in a form that clearly yields  Euler-Lagrange equations  that are second order in derivatives of the metric, as expected from Gauss-Bonnet theory.  

\section{Hamiltonian Analysis}

\label{hamiltonian}
We work with the dimensionally reduced two-geometry and write $g_{ab}$ in ADM form
\be
ds^2 = \erho \left( -\sigma^2 dt^2 + \left( dx + Ndt \right)^2 \right),
\label{adm}
\ee
where $\rho$, $\sigma$ and $N$ are functions of   $x$ and $t$. 
Since the theory is diffeomorphism invariant we know that  the total Hamiltonian will be of the form \cite{TZ}
\be
H = \int dx \left( \sigma \mathcal{G} + N \mathcal{F} \right)
\ee
The metric functions $\sigma$ and $N$ are Lagrange multipliers and $\mathcal{G}$ and $\mathcal{F}$ are the Hamiltonian and diffeomorphism constraints.  Moreover, as noted in \cite{JL97} the form of $\mathcal{F}$ is completely specified by the choice of parametrization and the fact that $\mathcal{F}$ generates spatial diffeomorphisms. Specifically, in the present case the gravitational part is:
\be
\mathcal{F}_G= \Pi_\rho \rho'-\Pi_\rho'+\Pi_\phi \phi'
\label{diffeo constraint}
\ee

We first focus on the gravitational sector of the theory. 
\subsection{Gravitational Hamiltonian}
The Hamiltonian analysis for vacuum 5-D Gauss-Bonnet theory was done in \cite{JL97} using the geometrodynamical formulation of Kuchar \cite{kuchar}. The algebra gets considerably more complicated in higher dimensions and as additional orders in derivatives are added in the more general case of Lovelock gravity. Here we take a simpler approach that is well adapted to the class of gauges we wish to consider.

First of all it is important to note that for the action $S_G$ only the term containing the Ricci scalar contains derivatives of the metric. Moreover, since the 2-D Ricci scalar for the parametrization (\ref{adm}) is
\be
R= - 2\tilde{\Box} \rho
\ee
where $\tilde{\Box}$ refers to the D'Alembertian of the metric:  $\left( -\sigma^2 dt^2 + \left( dx + Ndt \right)^2 \right)$, it follows that after an integration by parts, the action contains terms at most linear in the time derivative of $\rho$. This in turn implies that the gravitational part of the action can formally be written in the form:
\be
S_G=\int dtdx \left(B_1(\phid,\phi,\rho,\sigma,N)\rhod+B_0(\phid,\phi,\rho,\sigma,N)\right) \,\, .
\label{formal action}
\ee
Thus
\be
\Pi_\rho = B_1(\phid,\phi,\rho,\sigma,N)
\label{Pi_rho}
\ee
which in principle at least can be solved for $\phid$:
\be
\phid=\phid(\Pi_\rho,\rho,\phi,\sigma,N)
\ee
Note that in the above and what follows we do not specify separately dependence on spatial derivatives of a field separately from dependence on the field itself.

The Legendre transformation then yields a Hamiltonian of the form:
\be
H_G =  \int dx (\Pi_\phi\dot{\phi}(\Pi_\rho,\rho,\phi,\sigma,N) - B_0(\dot\phi(\Pi_\rho,\rho,\phi,\sigma,N),\rho,\phi,\sigma,N)).
\label{H2}
\ee
where as indicated $\phid$ is an implicit function of the other phase space variables as obtained from (\ref{Pi_rho}) but does {\bf not} depend on $\Pi_\phi$.
The crucial property of the above is that the $\dot{\rho}$ terms have cancelled and the only dependence on $\Pi_\phi$ is in the first term.

We now make the above explicit for the action $S_G=S_{EH}+S_{GB}$. In particular, we have:
\be
\label{generalaction}
S_G = \Theta \int dxdt (A_4 \phid^4 + A_3 \phid^3 + A_2 \phid^2 + A_1 \phid + A_0)
\ee
where $\Theta = 8(n-1)\nu^{(n)}/(16\pi G_D n)$,
\begin{align}
\label{A_0}
& A_0 = \frac{2}{\sigma} \left(\gamma \frac{n^2}{4} \phi^{1-4/n} + \frac{2n}{8(n-1)} \phi \right) (\phip N \rhod - \phip N^2 \rhop - \phip N \Np + \sigma^2 \phip \rhop + \sigma \sigp \phip) \\ \nonumber
& + \gamma \frac{1}{\sigma^3 e^{2\rho}} \Biggl[ 2\sigma^3\left( \frac{1}{3} \left(\frac{N}{\sigma}\right)^2 -1 \right)\frac{N}{\sigma}\frac{\phip^3}{\phi}\rhod \\ \nonumber
& + (\sigma^2 - N^2)^2 \left( \frac{\phip^4}{n \phi^2} - \frac{\sigp \phip^3}{\sigma \phi} - \frac{\phi^{\prime \prime} \phip^2}{\phi} \right) - \frac{\phip^3}{\phi} 2N(\sigma^2 - N^2) \left( \frac{N \sigp}{\sigma} -\Np \right) \Biggr] \\ \nonumber
& + \gamma \frac{(n-3)n^2}{2} \left( \frac{\phi^{-4/n} (\sigma^2 - N^2) }{n \sigma} \phip^2 + e^{2\rho} \sigma\frac{n}{8} \phi^{2-8/n} \right) + e^{2\rho} \sigma \frac{n^2}{8} \phi^{2-4/n} + \frac{(\sigma^2 - N^2)}{2 \sigma} \phip^2,
\end{align}
\begin{align}
\label{A_1}
& A_1 = \frac{-\gamma}{e^{2\rho}}\left(-\left( \frac{1}{3} \left(\frac{N}{\sigma}\right)^2 +1 \right)\frac{N}{\sigma}\frac{\phip^3}{\phi^2} -\frac{4\rhop\phip^2}{\phi}\frac{N}{\sigma} + \frac{2\phip^2}{\phi}\left( \frac{\Np}{\sigma} - \frac{N\sigp}{\sigma^2} \right) + \frac{4N\phi^{\prime \prime} \phip}{\sigma \phi} \right) \\ \nonumber
& + \frac{2}{\sigma} \left(\gamma \frac{n^2}{4} \phi^{1-4/n} + \frac{2n}{8(n-1)} \phi \right) (\rhop N + \Np - \rhod) + \gamma (n-3)n \frac{N}{\sigma} \phi^{-4/n} \phip + \frac{N}{\sigma} \phip \\ \nonumber
& + \gamma \frac{1}{\sigma^3 e^{2\rho}} \left[ \frac{2\phip^2}{\phi^2} (\sigma^2 - N^2) \left( \phi \rhod - \frac{N \phip}{2} + \rhop N - \frac{\sigp \phi N}{\sigma} + \frac{2 \phip N}{n} \right) - \frac{2N^2 \phip^2}{\phi} \left( \frac{N \sigp}{\sigma} - \Np \right) \right],
\end{align}
\begin{align}
\label{A_2}
& A_2 = \gamma \frac{1}{\sigma^3 e^{2 \rho}} \Biggl[ -2(\sigma^2 - N^2) \frac{\phip^2}{n \phi^2} -2(\sigma^2 + N^2) \frac{\rhop \phip}{\phi} -\frac{2(n-2)}{n} \frac{N^2 \phip^2}{\phi^2} + \frac{2 \sigma^2 \phi^{\prime \prime}}{\phi} - \frac{2 \phip N \Np}{\phi} \\ \nonumber
& +\frac{2N \rhod \phip}{\phi} \Biggr] -\gamma \frac{(n-3)n \phi^{-4/n}}{2 \sigma} - \frac{1}{2 \sigma},
\end{align}
\be
\label{A_3}
A_3 = \gamma \frac{1}{\sigma^3 e^{2 \rho}} \frac{2}{3\phi} \left[ \frac{(n-3)}{n} \frac{2N\phip}{\phi} - \rhod + \Np + N \rhop \right],
\ee
\be
\label{A_4}
A_4 = \gamma \frac{1}{\sigma^3 e^{2 \rho}} \frac{(n-3)}{n}\left( \frac{-1}{3 \phi^2} \right) 
\ee
Here $\gamma$ is defined as $\gamma = \beta \frac{16(n-1)(n-2)}{n^2}$ and the dots and primes represent differentiation with respect to the $t$ and $x$ coordinates respectively.

For completeness we note that the conjugate momentum of $\phi$ is:
\be
\Pi_\phi = A_1+ 2A_2\dot{\phi} + 3A_3\dot{\phi}^2+4A_4\dot{\phi}^3
\ee

\subsection{Partial Gauge Fixing}
\label{PGF}

The above general discussion suggests a particularly convenient family of gauges, namely those for which the areal radius is time independent. We therefore consider the class of gauge fixing conditions:
\be
\chi_f=\phi-f(x) \approx 0
\label{chi f}
\ee
If $f(x)=x^{2/n}$, then this gives $x=r$, i.e. the spatial coordinate is equal to the areal radius, but for the moment we will consider the more general case.

The $\approx$ denotes the fact that the gauge fixing condition can only be imposed on initial data. If it is a good gauge fixing condition then the requirement that it be preserved by the time evolution will yield a condition on one of the lagrange multipliers $\sigma$ or $N$. $\chi_f$ must be second class with some linear combination of the constraints. Clearly:
\be
\dot{\chi_f} = (\frac{\partial_f \chi}{\partial \phi}) \lbrace \phi, H \rbrace = \dot{\phi}=0
\ee
From Eq.(\ref{Pi_rho}) we can deduce that the required condition on the phase space variables is:
\be
\Pi_\rho = B_1(\phid=0,\phi,\rho,\sigma,N)
\label{Pi_rho gauge}
\ee
which determines the shift $N$ in terms of $\sigma$, $\rho$, $\Pi_\rho$ and $\phi$. The latter is no longer dynamical.

In order to set $\chi$ strongly equal to zero in the Hamiltonian (as opposed to the equations of motion) we must use the formalism of Dirac brackets.  Fortunately, since $\phi$ is no longer a phase space variable, the Poisson bracket of all phase space variables with $\chi$ is zero.  This means that the Dirac bracket is the same as the Poisson bracket.

The next step in the gauge fixing procedure is to use the diffeomorphism constraint to solve for $\Pi_\phi$ in terms of the other phase space variables. While this is indeed possible, it is not necessary in this case because the general form of the Hamiltonian 
(\ref{H2}) implies that when $\dot{\phi}=0$, $\Pi_\phi$ is not present in the Hamiltonian.  
Thus in this family of gauges, the gravitational part of the Hamiltonian is simply:
\be
\label{H2-2}
H_G =  -\int dx B_0.
\ee
where it is understood that in $B_0$, the shift $N$ is replaced by its value as determined by (\ref{Pi_rho gauge}).


To find the Hamiltonian with $\phid = 0$ we use (\ref{H2-2}), noting that $B_0(\phid = 0) = A_0(\rhod = 0)$.  This gives
\begin{align}
\label{H3}
& H_G = -\Theta \int dx \Biggl[ \frac{2}{\sigma} \left(\gamma \frac{n^2}{4} \phi^{1-4/n} + \frac{2n}{8(n-1)} \phi \right) (- \phip N^2 \rhop - \phip N \Np + \sigma^2 \phip \rhop + \sigma \sigp \phip) \\ \nonumber
& + \gamma \frac{1}{\sigma^3 e^{2\rho}} \left[ (\sigma^2 - N^2)^2 \left( \frac{\phip^4}{n \phi^2} - \frac{\sigp \phip^3}{\sigma \phi} - \frac{\phi^{\prime \prime} \phip^2}{\phi} \right) - \frac{\phip^3}{\phi} 2N(\sigma^2 - N^2) \left( \frac{N \sigp}{\sigma} -\Np \right) \right] \\ \nonumber
& + \gamma \frac{(n-3)n^2}{2} \left( \frac{\phi^{-4/n} (\sigma^2 - N^2) }{n \sigma} \phip^2 + e^{2\rho} \sigma \frac{n}{8} \phi^{2-8/n} \right) + e^{2\rho} \sigma \frac{n^2}{8} \phi^{2-4/n} + \frac{(\sigma^2 - N^2)}{2 \sigma} \phip^2 \Biggr] 
\end{align}
where $\Pir$ is calculated to be
\be
\label{rho1}
\Pir =2 \Theta  \left(\gamma \frac{n^2}{4} \phi^{1-4/n} + \frac{2n}{8(n-1)} \phi \right) \phip \frac{N}{\sigma} + \Theta \frac{2\gamma}{e^{2\rho}}\left( \frac{1}{3} \left(\frac{N}{\sigma}\right)^2 -1 \right)\frac{N}{\sigma}\frac{\phip^3}{\phi} 
\ee
which is a cubic equation that can be solved for $N/\sigma$.
As noted in \cite{JL97} there can in principle be multiple real roots, depending on the parameters involved, which would yield different dynamical spacetimes. In that work, the boundary conditions effectively resolved this potential ambiguity. In the appendix we show that in our case as well there is a unique real branch, and hence unique dynamics, for the flat slice, generalized P-G coordinates that we consider.

We therefore assume that  there is a unique solution that specifies implicitly $N/\sigma$ as a function of the remaining phase space variables and we use this to write the Hamiltonian in terms of one Lagrange multiplier, $\sigma$, and one constraint 
\begin{align}
\label{H6}
& H_G = -\Theta \int dx \sigma \erho \Biggl[ -\gxx^\prime \phip {\cal D},_\phi - 2\gxx\phi^{\prime \prime} {\cal D},_\phi - 2 \gxx {\cal D},_{\phi \phi} \phip^2 \\ \nonumber
& \frac{\gamma \phip^3 (\gxx^2)^\prime}{2 \phi} + \gamma \gxx^2 \left( \frac{2 \phip^2 \phi^{\prime \prime}}{\phi} - \frac{(n-1)\phip^4}{n \phi^2} \right)
\\ \nonumber & + \left( \gamma (n-3)n \phi^{-4/n} +1 \right) \gxx \phip^2 / 2 + \gamma(n-3)n^3 \phi^{2-8/n} / 16 + n^2 \phi^{2-4/n} / 8 \Biggr]
\end{align}
where ${\cal D}$ is the coefficient of $R$ in the action, equation (\ref{action}) (${\cal D},_\phi = \gamma \frac{n^2}{4} \phi^{1-4/n} + \frac{2n}{8(n-1)} \phi$) and $g^{11}$ is given by
\be
\label{gxx1}
\gxx = \frac{1}{\erho} \left( 1 - \left( \frac{N}{\sigma} \right)^2 \right)
\ee

After some algebra, the Hamiltonian can be written as
\be
\label{H7}
H_G = \int dx \sigma \left( \frac{-n \erho }{2 \phi^{2/n-1} \phip} \right) \m^\prime,
\ee
where we have defined the mass function $\m$ \cite{HM06b} as:
\begin{align}
\label{mass1}
& \frac{\m}{\Theta} = \\ \nonumber
& \frac{n}{8(n-1)} \left[ n \phi^{2-2/n} \left( 1 - \frac{4 \gxx \phi^{4/n-2} \phip^2}{n^2} \right) + \frac{\gamma n^2 (n-1) \phi^{2-6/n}}{2} \left( 1 - \frac{4 \gxx \phi^{4/n-2} \phip^2}{n^2} \right)^2 \right],
\end{align}

Using:
\be
|\nabla r|^2= g^{11} (r')^2 = g^{11}\frac{n^2}{4} \phi^{2-4/n}(\phi')^2
\ee
one can verify that $\m$  agrees with the mass function defined in equation (2.7) of \cite{HM08}. It is therefore the EGB generalization of the Misner-Sharp mass function. It is invariant under coordinate transformations that preserve spherical symmetry and asymptotes to the ADM mass for asymptotically flat spacetimes. Clearly $\m$ is constant for vacuum spacetimes.

Eq.(\ref{mass1}) is a quadratic that can be inverted to express $\gxx$ and (via (\ref{gxx1})) $N/\sigma$ in terms of the mass function:
\begin{align}
\label{gxx}
& \gxx = \frac{1}{\erho} \left( 1 - \left( \frac{N}{\sigma} \right)^2 \right) = \nn
& \frac{n^2 \phi^2}{4 \phi^{4/n} \phip^2} - \frac{n \phi^2}{4 \gamma (n-1) \phip^2} \left( -1 \pm \sqrt{1 + 16 \gamma (n-1)^2 \phi^{-2-2/n} \m /n\Theta} \right)
\end{align}
From the above, we see that in vacuum, when $\m=M$ is constant, the above yields the two branches of the well-known vacuum solution of EGB theory. For the following, we choose the plus-branch in (\ref{gxx}) in order that our solutions
reduce to those in Einstein gravity in the limit $\gamma\to 0$. However, nothing in principle precludes the other choice, which would have interesting consequences for the end-point of gravitational collapse.

This in turn puts the ADM metric in the form:
\be
ds^2 = \erho \left( -\sigma^2 \erho \gxx dt^2 \pm 2 \sigma \sqrt{1-\erho \gxx} dtdr + dr^2 \right).
\label{adm metric 2}
\ee
with $\gxx$ given in (\ref{gxx}) above. Thus, the partially gauge fixed geometry is now implicitly specified in terms of $\rho,\Pi_\rho$ and the the lapse $\sigma$. 
 
Finally, noting that the condition for the existence of an apparent horizon is that the null expansion vanishes:
\be
|\nabla r|^2=g^{11}(r')^2=0
\ee
it is easy to verify that by choosing the + (-) in front of the off-diagonal term (\ref{adm metric 2})one can guarantee that the slicings are regular across future (past) horizons. The exception to this is the gauge choice $N=0$ (Schwarzschild gauge) which breaks down at all horizons.

\subsection{Adding the Matter Contribution}
In order to add the scalar field contribution to the action we dimensionally reduce equation (\ref{scalaraction}) to get

\begin{align}
\label{scalaraction2}
& S_M = -\frac{\nu^{(n)}}{2}\int dxdt \sqrt{-g} \phi^2 |\nabla \psi|^2 = \\ \nonumber
& -\frac{\nu^{(n)}}{2}\int dxdt \frac{\phi^2}{\sigma} \left( -\psid^2 + 2N\psip \psid + (\sigma^2 - N^2) \psip^2 \right),
\end{align}
which gives the momentum conjugate to $\psi$,

\be
\label{Pip}
\Pis = \frac{\nu^{(n)} \phi^2}{\sigma} \left( \psid - N \psip \right).
\ee
From this we can write the matter action as

\begin{align}
\label{scalaraction3}
& S_M = \int dxdt \psid \Pis - \int dxdt \left[ \frac{\nu^{(n)} \phi^2}{2\sigma} \left( \frac{\sigma \Pis}{\nu^{(n)} \phi^2} +N\psip \right)^2 + \frac{\sigma \phi^2 \nu^{(n)}}{2} \left(1 - \sub^2 \right) \psip^2 \right] = \\ \nonumber
& \int dxdt \psid \Pis - \int dxdt \sigma \left[ \frac{1}{2} \left( \frac{\Pis^2}{\nu^{(n)} \phi^2} + \nu^{(n)} \phi^2 \psip^2 \right) + \left( \frac{N}{\sigma} \right) \Pis \psip \right].
\end{align}

From equation (\ref{scalaraction3}) it is clear that the total partially gauge-fixed Hamiltonian is given by

\be
\label{H8}
H = \int dx \sigma \left[ \frac{-n \erho }{2 \phi^{2/n-1} \phip} \m^\prime + \frac{1}{2} \left( \frac{\Pis^2}{\nu^{(n)} \phi^2} 
+ \nu^{(n)} \phi^2 \psip^2 \right) + \left( \frac{N}{\sigma} \right) \Pis \psip \right]
\ee
where $\left( \frac{N}{\sigma} \right)$ is understood to the be real branch of the solution to (\ref{rho1}). 
The Hamiltonian (\ref{H8}) is the generalization to G-B gravity of the partially reduced Hamiltonian derived in \cite{DGK06}. Note that it is valid for any gauge fixing condition $\chi_f$ in the form (\ref{chi f}). For concreteness we henceforth specify:
\be
f(x)= x^{n/2}
\label{f}
\ee 
which chooses the areal radius as the spatial coordinate: $x=r=\phi^{2/n}$. 

At this stage, the dynamical system consists of four phase space variables $(\rho,\Pi_\rho,\psi,\Pi_\psi)$ and there is a single first class constraint, namely the Hamiltonian constraint. We now look at the complete reduction to only the physical degrees of freedom associated with the scalar field.

\section{Completely Reduced Theory}
\label{gauge}

For our second gauge choice we follow \cite{JZ09} and choose
\be
\label{gauge2}
\xi = \erho - 1\approx 0.
\ee
It is clear from (\ref{adm}) and (\ref{4 metric}) that this choices yields spatial slices that are flat. It therefore generalizes P-G coordinates. To further understand the significance of this choice, note that
using (\ref{gxx}) in (\ref{adm metric 2}) and setting $\chi$ to zero in the limit as $\gamma \to 0$ (choosing the plus sign), we get $g^{11} \to 1-2G\m/r$.  Inserting this into (\ref{adm metric 2}) and setting $\xi$ to zero we find that the metric becomes
\be
ds^2 = -\sigma^2 \left( 1-\frac{2G\m}{r} \right)  dt^2 + 2 \sigma \sqrt{\frac{2G\m}{r}} dtdr + dr^2.
\ee
This shows that the gauge choice $\xi = 0$ puts the metric into the Gauss-Bonnet, non-static generalization of the Painleve Gullstrand co-ordinates.  Using $\xi = 0$ also allows us to write the consistency condition for the first gauge choice, equation (\ref{gxx}), as
\be
\label{Nsig}
\frac{N}{\sigma} = \sqrt{\frac{\phi^{4/n}}{\gamma n(n-1)} \left( -1 + \sqrt{1 + 16 \gamma (n-1)^2 \phi^{-2-2/n} \m /n\Theta} \right)}.
\ee

To ensure that the second gauge condition is conserved in time we must insist that $(d/dt)(\erho - 1) = \{(\erho - 1), H\} = 0 \rightarrow (\delta/\delta \Pir)H = 0$.  From (\ref{rho1}) we know that the $\Pir$ dependence in the Hamiltonian is in the $(N/\sigma)$ terms.  So we can write
\begin{align}
\label{H10}
\frac{\delta H}{\delta \Pir} & = \frac{\delta}{\delta \Pir} \int dx \left[ \sigma^\prime \m + \sigma \left( \frac{N}{\sigma} \right) \Pis \psip \right] \\ \nonumber
& = \sigma^\prime \frac{\partial \m}{\partial (N/\sigma)} \frac{\partial (N/\sigma)}{\partial \Pir} + \sigma \frac{\partial (N/\sigma)}{\partial \Pir} \Pis \psip
\end{align}
from which we can write the consistency condition
\begin{align}
\label{sigma1}
& \sigma^\prime \frac{\partial \m}{\partial (N/\sigma)} + \sigma \Pis \psip = \\ \nonumber
& \sigma^\prime \frac{n}{8(n-1)} \left( 2n\Theta\phi^{2-2/n} (N/\sigma) + 2\gamma\Theta n^2 (n-1) \phi^{2-6/n} (N/\sigma)^3  \right) + \sigma \Pis \psip = 0,
\end{align}
where it is understood that we use (\ref{Nsig}) to write $(N/\sigma)$ in terms of the mass function $\m$.

Using $\chi = \xi = 0$ we can now write down the fully gauge fixed Hamiltonian as
\be
\label{H9}
H = \int dx \sigma \left[-\m^\prime + \frac{1}{2} \left( \frac{\Pis^2}{\nu^{(n)} \phi^2} + \nu^{(n)} \phi^2 \psip^2 \right) + \left( \frac{N}{\sigma} \right) \Pis \psip \right].
\ee


Using Hamilton's equations we find
\be
\label{psidot}
\dot{\psi} = \sigma \left( \frac{\Pis}{\nu^{(n)} \phi^2} + \left( \frac{N}{\sigma} \right) \psip \right)
\ee
\be
\label{Pisdot}
\dot{\Pi}_\psi = \left[ \sigma \left( \nu^{(n)} \phi^2 \psip + \left( \frac{N}{\sigma} \right) \Pis \right) \right]^\prime
\ee
These equations, along with the consistency conditions (\ref{Nsig}) and (\ref{sigma1}) and the Hamiltonian constraint
\be
\label{Cconstraint}
-\m^\prime + \frac{1}{2} \left( \frac{\Pis^2}{\nu^{(n)} \phi^2} + \nu^{(n)} \phi^2 \psip^2 \right) + \left( \frac{N}{\sigma} \right) \Pis \psip = 0
\ee
determine the evolution of a collapsing scalar field. 

\section{Conclusion}
\label{conclusion}

Lovelock gravity has become increasingly popularized in recent years, as other theories have required higher dimensional analogues to General Relativity.   We have presented here the Hamiltonian formulation  of the lowest order non-trivial Lovelock theory, Gauss-Bonnet gravity, minimally coupled to a scalar field in the spherically symmetric case. 

The reduced Hamiltonian system (\ref{sigma1} - \ref{Cconstraint}) is well suited for the study of gravitational collapse of a scalar field along with the resultant critical phenomena.  Our calculations utilize a gauge choice that selects non-static Painlev\'{e}-Gullstrand coordinates, a coordinate system that is regular across the horizon during black hole formation. Moreover, this system is easily generalized to any matter field whose action does not contain derivatives of the metric.

Future work will involve numerically studying gravitational collapse of a scalar field in this scenario to see how 
non-linearity in curvature modifies the well-known critical phenomena \cite{collapse} in Einstein gravity.

\section{Acknowledgements}
This work was supported in part by the Natural Sciences and Engineering Council of Canada. GK is grateful to Jorma Louko and Hideki Maeda for useful conversations.  TT would like to thank the University of Manitoba for funding.

\section{Appendix}
We now solve equation \ref{rho1} for $N/\sigma$.  Equation \ref{rho1} can be written in the form
\be
\label{rho2}
C_3 \sub^3 + C_1 \sub -\Pir = 0
\ee
where
\be
C_3 = \frac{\Theta2\gamma \phip^3}{3 e^{2\rho} \phi} \quad and \quad C_1 = \Theta \left( 2\phip(\gamma \frac{n^2}{4} \phi^{1-4/n} + \frac{2n}{8(n-1)} \phi) - \frac{2 \gamma \phip^3}{\phi e^{2 \rho}} \right)
\ee

Providing $C_1/C_3>0$,  there is only one real solution, namely:
\be
\sub = \frac{\left(108 \Pir + 12 \sqrt{3} C_3^6\sqrt{4C_1^3/C_3 + 27 \Pir^2} \right)^{1/3}}{6C_3} - \frac{2C_1}{\left(108 \Pir + 12 \sqrt{3} C_3^6\sqrt{4C_1^3/C_3 + 27 \Pir^2} \right)^{1/3}}
\ee
The condition that $C_1/C_3>0$ can be simplified to 
\be
\frac{e^{2\rho}}{4\gamma\phip^2}\left( \gamma n^2 \phi^{2(n-2)/n} + \frac{n\phi^2}{n-1} \right) - 1 > 0
\ee
for the solution to equation \ref{rho2} to be real.  This won't be identically true so to proceed we must be careful with our gauge choices or possibly our allowed range of $\gamma$.  For the flat slice coordinates of Section (\ref{gauge}) $x=r$ (ie $\phip=(n/2)\phi^{1-2/n}$) and $\erho=1$ so that the condition is in fact identically satisfied.

\end{document}